\documentclass[useAMS,usenatbib,usegraphicx]{mn2e}
\usepackage{bm}
\usepackage{times}

\def\tff{t_{\rm ff}}
\newcommand{\hii}{{H}\,\textsc{ii}}

\title[SMA observation of II Zw 40]
{Central free--free-dominated 880 $\micron$ emission in
II Zw 40}
\author[Hirashita]{Hiroyuki Hirashita$^1$\thanks{E-mail:
    hirashita@asiaa.sinica.edu.tw}\\
$^1$Institute of Astronomy and Astrophysics, Academia Sinica,
P.O. Box 23-141, Taipei 10617, Taiwan
}
\date{2011 July 29}

\pagerange{\pageref{firstpage}--\pageref{lastpage}} \pubyear{2011}

\begin{document}
\label{firstpage}
\maketitle

\begin{abstract}
The central star-forming region in a blue compact
dwarf galaxy, II Zw 40, was observed in the
340~GHz ($880~\micron$) band at $\sim 5$ arcsec
(250 pc) resolution with the Submillimetre Array
(SMA). A source associated with the central
star-forming complex was detected with a flux of
$13.6\pm 2.0$ mJy. The structure is more extended
than the beam in the east--west direction. The SMA
880 $\micron$ flux
is analyzed by using theoretical models of radio
spectral energy distribution along with
centimetre interferometric measurements in the
literature.
{We find (i) that the SMA 880 $\micron$
flux is dominated
($\sim 75$ per cent) by free--free emission from the
central compact
star-forming region, and (ii) that
the contribution from dust emission
to the SMA 880 $\micron$ flux is at most
$4\pm 2.5$ mJy. We also utilize our models to
derive the radio--FIR relation of the II Zw 40 centre,
suggesting that free--free absorption at low
frequencies ($\nu\la$ several GHz; $\lambda\ga$ several cm)
and spatial extent of dust affect
the radio--FIR relation.}
\end{abstract}

\begin{keywords}
dust, extinction --- galaxies: dwarf ---
galaxies: evolution ---galaxies: individual (II Zw 40)
--- H \textsc{ii} regions --- submillimetre: galaxies
\end{keywords}

\section{Introduction}

Blue compact dwarf galaxies (BCDs) generally host
compact and ongoing star formation activities in
metal-poor and gas-rich environments
\citep*{sargent70,vanzee98}. Because both low
metallicity and rich gas content indicate an early
evolutionary stage, BCDs can be used as nearby
laboratories of primeval galaxies which should exist
at high redshift. In some BCDs, the most active class
of star formation is taking
place in super star clusters (SSCs)
{\citep{turner98,kobulnicky99,johnson03},
which are expected to `mimic' the
starburst in high-redshift primeval galaxies}.

II Zw 40 is a well studied BCD with a low
oxygen abundance,
$12+\log (\mathrm{O/H})=8.13$ \citep{thuan05}.
This galaxy hosts a high star formation activity
associated with the SSCs in the centre. At
centimetre wavelengths, the central star-forming
region in II~Zw~40 is compact and optically
thick for free--free absorption, and is categorized
as a `supernebula' or `ultradense \hii\ region'
\citep{turner98,kobulnicky99}. Such a dense and
compact star formation activity is called `active'
mode in \citet{hunt03} and \citet{hirashita04}.
II Zw 40 is also classified
as a Wolf-Rayet galaxy: the Wolf-Rayet feature
indicates that the
typical age of the current starburst is a few Myr
\citep{vacca92}. \citet{buckalew05} derived an age of
2.6 Myr from the H$\alpha$ and H$\beta$ equivalent
widths. The Br$\gamma$ equivalent width also shows
age $\la 3$ Myr \citep{vanzi08}. Stellar spectral
synthetic models support young ages $\sim 2$ Myr,
although there is an underlying old stellar
population \citep{westera04}.

In order to trace dense `embedded' star-formation
activities, optically thin star formation indicators
are useful, such as far-infrared (FIR) dust luminosity
\citep[e.g.][]{kennicutt98,inoue00} and radio
(thermal plus non-thermal) luminosity \citep{condon92}.
Indeed, there is a correlation between
FIR and radio luminosities
in nearby star-forming galaxies
\citep[e.g.,][]{dejong85,helou85},
which is naturally explained if
both luminosities are strongly connected
with star formation activities \citep{volk89}.
However, the studies of radio--FIR relation are
biased to objects with
FIR detection \citep[e.g.][]{condon92}, which means
that a significant dust enrichment has already occurred.
Therefore, the evolution of
radio--FIR relation in
young primeval galaxies is not yet clear.
The gas
density, the magnetic field strength, and the
energy density of cosmic ray electrons
affect the evolution of radio emission
on various time-scales
{\citep{helou93,niklas97,murphy09,lacki10}},
while
the dust enrichment plays a role to
increase the FIR luminosity
\citep{hirashita_hunt08}.

There have been some observational studies on the
evolution of radio--FIR relation along the cosmic
age. The radio--FIR relation at moderate and high
redshifts ($z\la 4$) is broadly similar to that
at the local Universe
\citep*[e.g.][]{garrett02,gruppioni03,ibar08,
murphy09,michalowski10},
although there is also a slight indication of
evolution
\citep{vlahakis07,seymour09,michalowski10}.
The disadvantage of high-$z$ observations is
the difficulty in deriving quantities related to
galaxy evolution (age, metallicity, etc.)
with high accuracy. On the other hand, studies
of nearby metal-poor dwarf galaxies as
`laboratories' of primeval galaxies
provide an alternative way to approach the
properties of primeval galaxies.
\citet*{hopkins02} and \citet{wu08} conclude
that the radio--FIR relation of star-forming
dwarf galaxies is similar to that of normal
galaxies in spite of the difference
in metallicity \citep*[see also][]{klein91},
although \citet{cannon05,cannon06} find a
significant deviation from the canonical
radio--FIR relation for some
individual dwarf galaxies.

In this paper, we do not take a statistical
way, but investigate a single object in details.
II Zw 40 is suitable for studying radio--FIR
emission in a metal-poor young object, since
as mentioned above it hosts an extremely young
($\la 3$ Myr) active star formation with a
metallicity of 1/4 Z$_{\sun}$ (for the solar
metallicity, we adopt $12+\log\mathrm{O/H})=8.69$;
\citealt{lodders03}). In order
to spot the young star-forming component, a high
resolution is necessary. The star-forming
region is resolved well at centimetre
wavelengths, while
there is no information on FIR--submillimetre
(submm) dust emission on
such a small scale. A submm interferometric
observation is desired for the purpose of
resolving the
dust emission in the star-forming region.
Therefore, we observed II Zw 40 by the
Submillimetre Array (SMA; \citealt{ho04}), and
we report on this observation in this paper.
This observation, combined with centimetre radio
interferometric data in the literature, will enable
us to obtain the radio--FIR emission
properties (or radio--FIR relation) in the very young
star-forming region with a low metallicity.

This paper is organized as follows. We explain the
observations and the data reduction in
Section \ref{sec:obs}, and describe observational
results in Section \ref{sec:result}. In
Section \ref{sec:model}, we interpret the results
along with the radio data in the literature by using
theoretical models. In Section \ref{sec:firsed},
we analyze the contribution from dust emission to
the SMA flux.
We discuss the radio--FIR relation based on the
SMA observation in Section \ref{sec:discuss}.
Finally, we conclude in Section \ref{sec:conclusion}.
The distance to II Zw 40 is assumed to be $D=10.5$ Mpc
($cz=789$ km s$^{-1}$ with $H_0=75$ km s$^{-1}$ Mpc$^{-1}$).
At this distance, 1 arcsec corresponds to 50.9 pc.

\section{Observations and data}\label{sec:obs}

The SMA observation of II Zw 40 was carried out in the
340 GHz (880 $\micron$) band on 2010 March 30 in the
subcompact configuration. Six antennas were used with
projected antenna separations between 9.5 and 25 m. The
receivers have two sidebands, the lower and upper
sidebands, which covered the frequency ranges from 331.0
to 335.0 GHz,
and from 342.9 to 346.9 GHz, respectively. The visibility
data were calibrated with the MIR package, with Mars as
a flux calibrator (with an adopted flux of 1194 Jy),
quasars J0423$-$013 and J0730$-$116 as amplitude and
phase calibrators, and quasar 3C273 as a band pass
calibrator. The calibrated visibility data were imaged
and CLEANed with the MIRIAD package. The synthesized
beam has a size of FWHM $5.2''\times 4.4''$
($265~\mathrm{pc}\times 224~\mathrm{pc}$)
with a major axis position angle of $\sim 60\degr$.
The beam roughly matches the size of the region of
active star formation in the \textit{Hubble
Space Telescope} image \citep{calzetti07} and the
Very Large Array (VLA) image \citep{ulvestad07}.
The largest angular scale sampled by this observation
is $19''$.

\begin{table}
\centering
\begin{minipage}{80mm}
\caption{Radio data used in this paper.}
\label{tab:data}
\begin{tabular}{cccccc}
\hline
$\nu$ & $\theta_\mathrm{max}\,^\mathrm{a}$ & Flux &
Image\,$^\mathrm{b}$ & Fit\,$^\mathrm{c}$ &
Ref.\,$^\mathrm{d}$ \\
(GHz) & (arcsec) & (mJy) & & \\
\hline
1.4  & ---  & $30.0\pm 0.5$  & S  &  & 1 \\
1.5  & ---  & $29.9\pm 0.7$  & S  &  & 2\\
1.5  & ---  & $30.5\pm 1.5$  & S  &  & 3\\
4.8  & ---  & $22\pm 3$  & S   &  & 4 \\
5.0  & ---  & $22\pm 4$  & S  &  & 1 \\
5.0  & 10 & $15\pm 1$  & I  & & 5 \\
5.0  & 4  & $9\pm 1.5$  & M & $\surd$ & 5\\
5.0  & ---  & $21.0\pm 1.9$ & S & & 2 \\
8.3  & 7  & $12\pm 1$  & I & & 5 \\
8.3  & 4  & $10\pm 1.5$ & M & $\surd$ & 5\\
15  & 4  & $14\pm 1.5$ & I & $\surd$ & 5\\
25 & --- & $18\pm 4$  & S  &  & 4\\
\hline
\end{tabular}

\medskip

$^\mathrm{a}$ The maximum size scale that is well sampled
by the interferometric observation.\\
$^\mathrm{b}$ `S', `I', and `M' represent single dish flux,
interferometric flux, and matched beam flux, respectively.\\
$^\mathrm{c}$ If the data is used to fit the radio emission
models
in Section \ref{sec:model}, this column is marked with
$\surd$.\\
$^\mathrm{d}$ References. 1) \citet*{jaffe78};
2) \citet{klein91}; 3) \citet{deeg93}; 4) \citet*{klein84};
5) \citet{beck02}.\\
\end{minipage}
\end{table}

\begin{table}
\centering
\begin{minipage}{55mm}
\caption{FIR and submm data used in this paper.}
\label{tab:fir}
\begin{tabular}{cccccc}
\hline
$\lambda$ & Flux & Image\,$^\mathrm{a}$ &
Ref.\,$^\mathrm{b}$ \\
($\micron$) & (mJy) & & \\
\hline
60  & $6610\pm 700$ & S & 1\\
65  & $6900\pm 700$ & S & 2\\
70  & $5580\pm 280$ & S & 3\\
90  & $6600\pm 700$ & S & 2\\
100 & $5800\pm900$ & S & 1\\
140 & $3700\pm 700$ & S & 2\\
160 & $3400\pm 900$ & S & 2\\
160 & $3140\pm 430$ & S & 3\\
450 & $400\pm 90$  & S  & 1 \\
450 & $248\pm 81$  & S  & 4 \\
850 & $90\pm 10$  & S  & 1 \\
850 & $98\pm 14$  & S & 4\\
880& $13.6\pm 2.0$ & I & 5\\
\hline
\end{tabular}

\medskip

$^\mathrm{a}$ `S' and `I' represent single dish flux and
interferometric flux, respectively.\\
$^\mathrm{b}$ References. 1) \citet{hunt05};
2) \citet{hirashita08}; 3) \citet{engelbracht08};
4) \citet{galliano05}; 5) this paper.
\end{minipage}
\end{table}

In order to estimate the contribution of free--free
emission to the 880 $\micron$ SMA flux, we use
radio data in the literature (Table~\ref{tab:data}).
The single-dish observations measure the fluxes from
the entire galaxy, while the interferometric data are
sensitive to angular sizes
smaller than $\theta_\mathrm{max}$. We also
adopt the `matched beam' VLA fluxes obtained with
$(u,\, v)$ data restricted to baselines greater than
20k$\lambda$ (i.e.\ sensitive to structures smaller than
4 arcsec) \citep{beck02}. These matched fluxes are used to
constrain the properties of the central active
star formation.
The discrepancy between the matched data and
the single-dish data is larger at lower frequencies,
which implies that the spectral slope is steeper
in the diffuse medium than in the central
star-forming region.

For the information on dust emission, we also
use FIR data in the literature as listed in
Table \ref{tab:fir}, where the fluxes except
for our SMA data are
all measured by single-dish observations and
represent those of the entire galaxy.
{We
only consider the `large grain' component
which is in radiative equilibrium with the
ambient interstellar radiation field
(see Section \ref{subsec:dust}).}
Thus,
we do not use the data at mid-infrared
and shorter wavelengths, where very
small grains and PAHs dominate dust
emission
\citep[e.g.][]{draine01,dopita06}
and the spectral energy distribution
(SED) is sensitive to
grain size distribution.

\section{Results}\label{sec:result}

\begin{figure}
\includegraphics[width=0.47\textwidth]{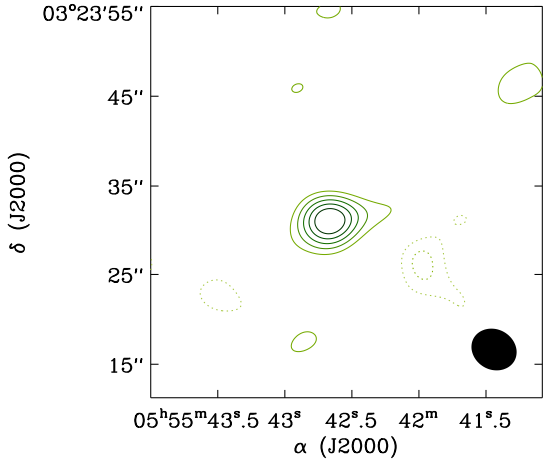}
\caption{SMA 880 $\micron$ continuum image of II Zw 40.
Solid contours are 2, 3, 4, 5, and 6 $\sigma$
(1 $\sigma =1.15$ mJy beam$^{-1}$), while dotted
contours are $-2$ and $-3$ $\sigma$. The beam is
shown in the lower right corner.\label{fig:image}}
\end{figure}

Figure \ref{fig:image} shows the obtained image.
The source is clearly detected with a peak intensity of
7.4 $\sigma$ (1 $\sigma =1.15$ mJy beam$^{-1}$)
{and} a total flux of
$13.6\pm 2.0$ mJy.
There is an extended feature in the east--west direction
with FWHM = 7.1 arcsec (the width of the 2 $\sigma$
contour is $\sim 10$ arcsec).

\begin{figure*}
\includegraphics[width=0.45\textwidth]{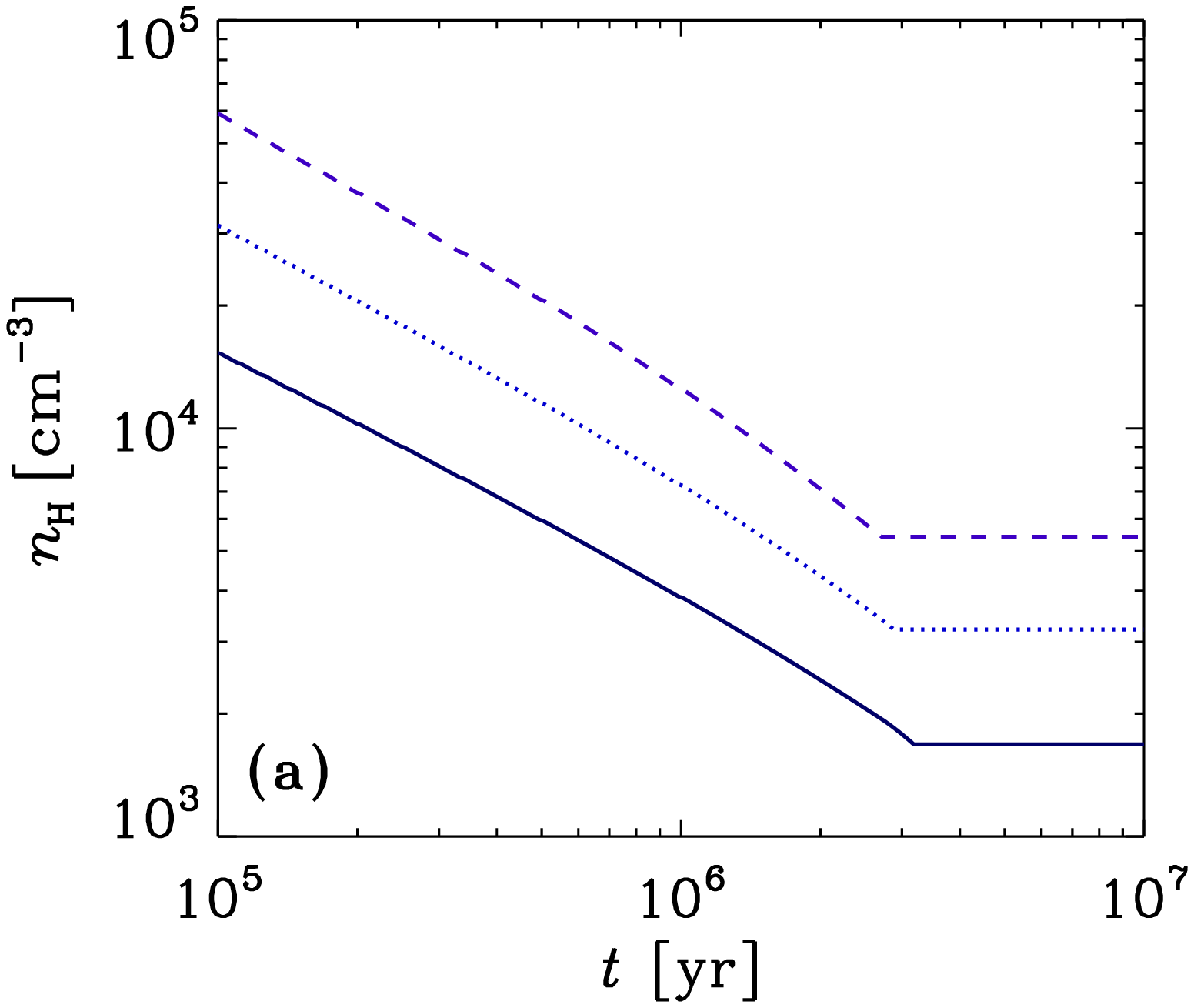}
\includegraphics[width=0.45\textwidth]{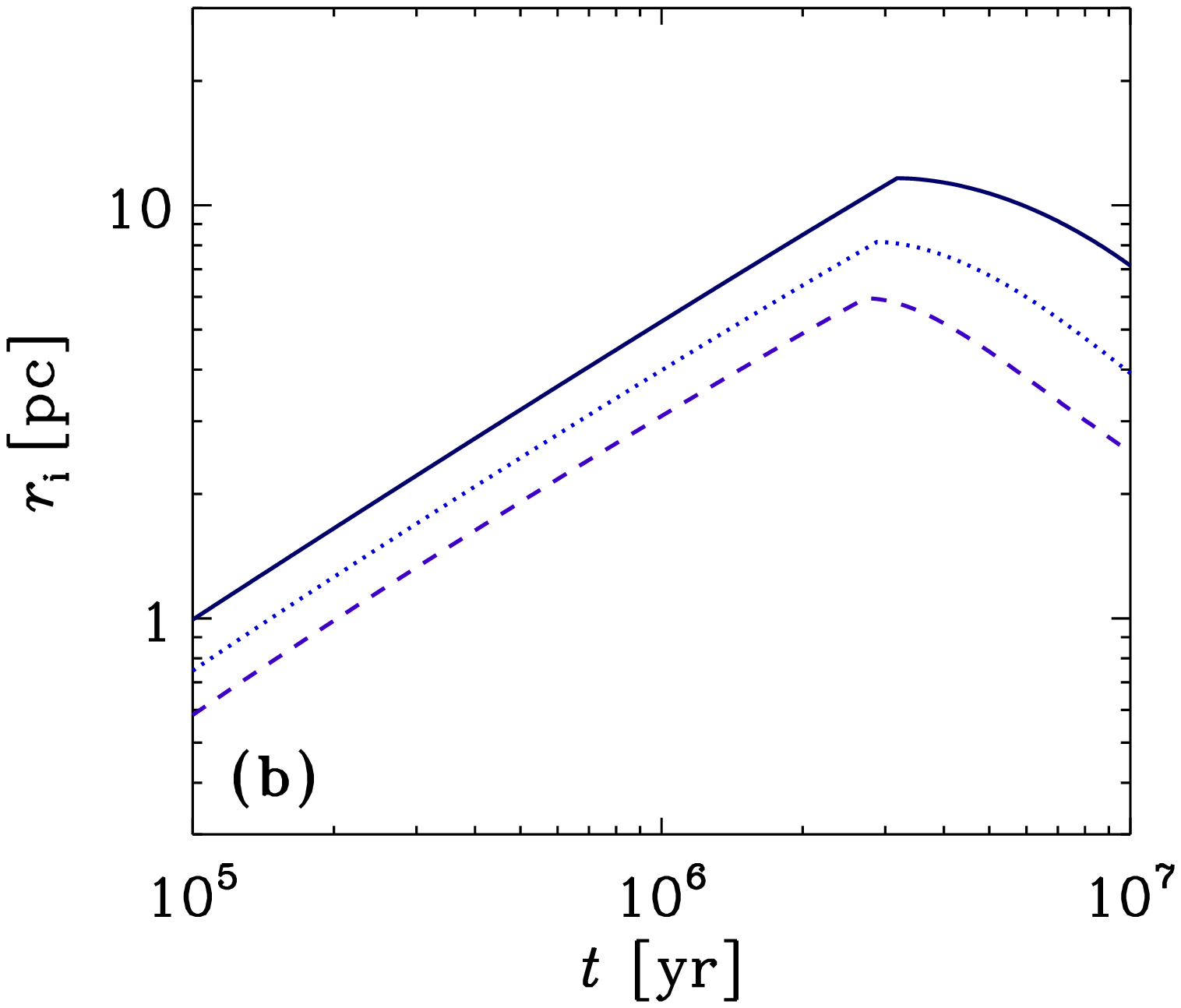}
\includegraphics[width=0.45\textwidth]{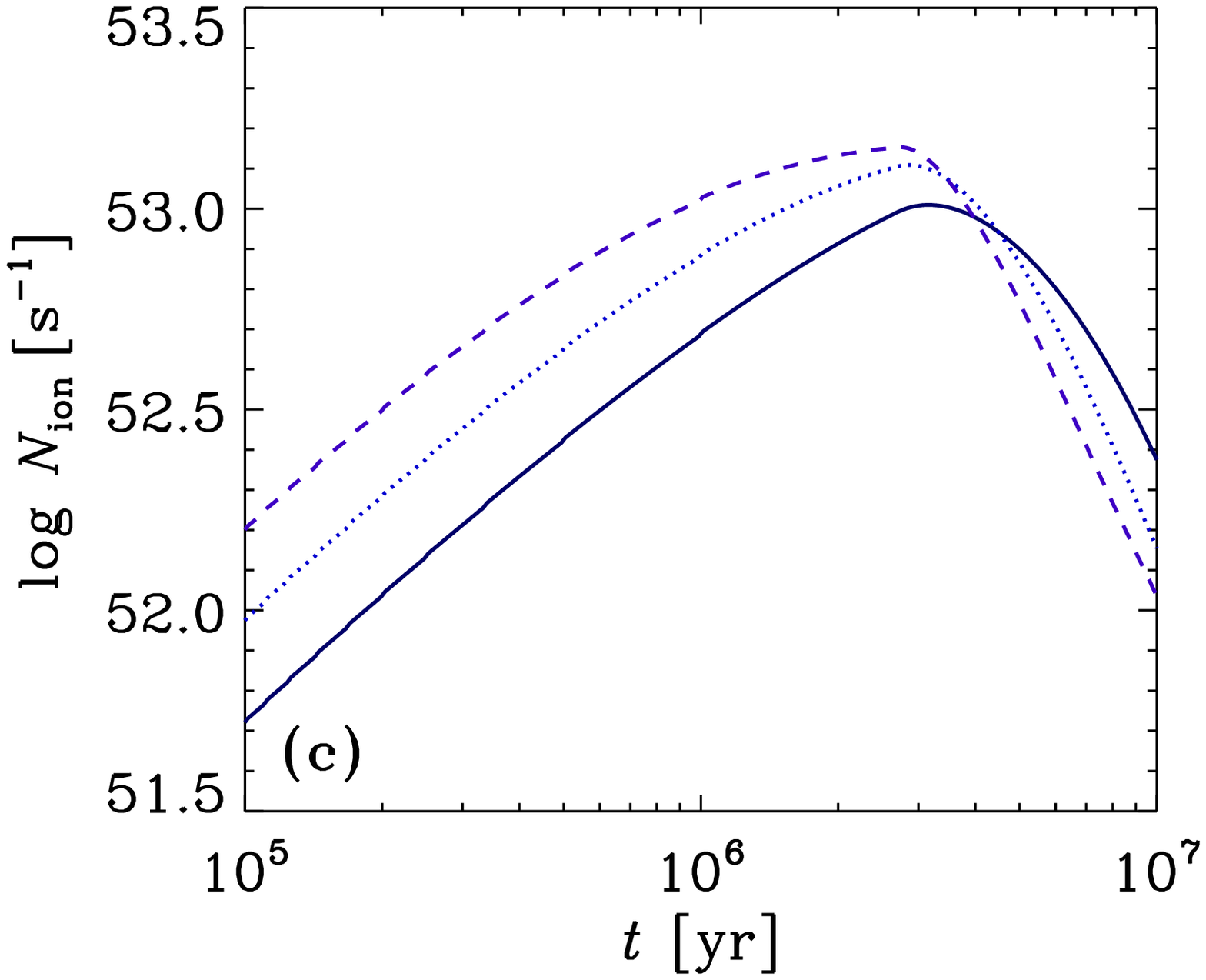}
\includegraphics[width=0.45\textwidth]{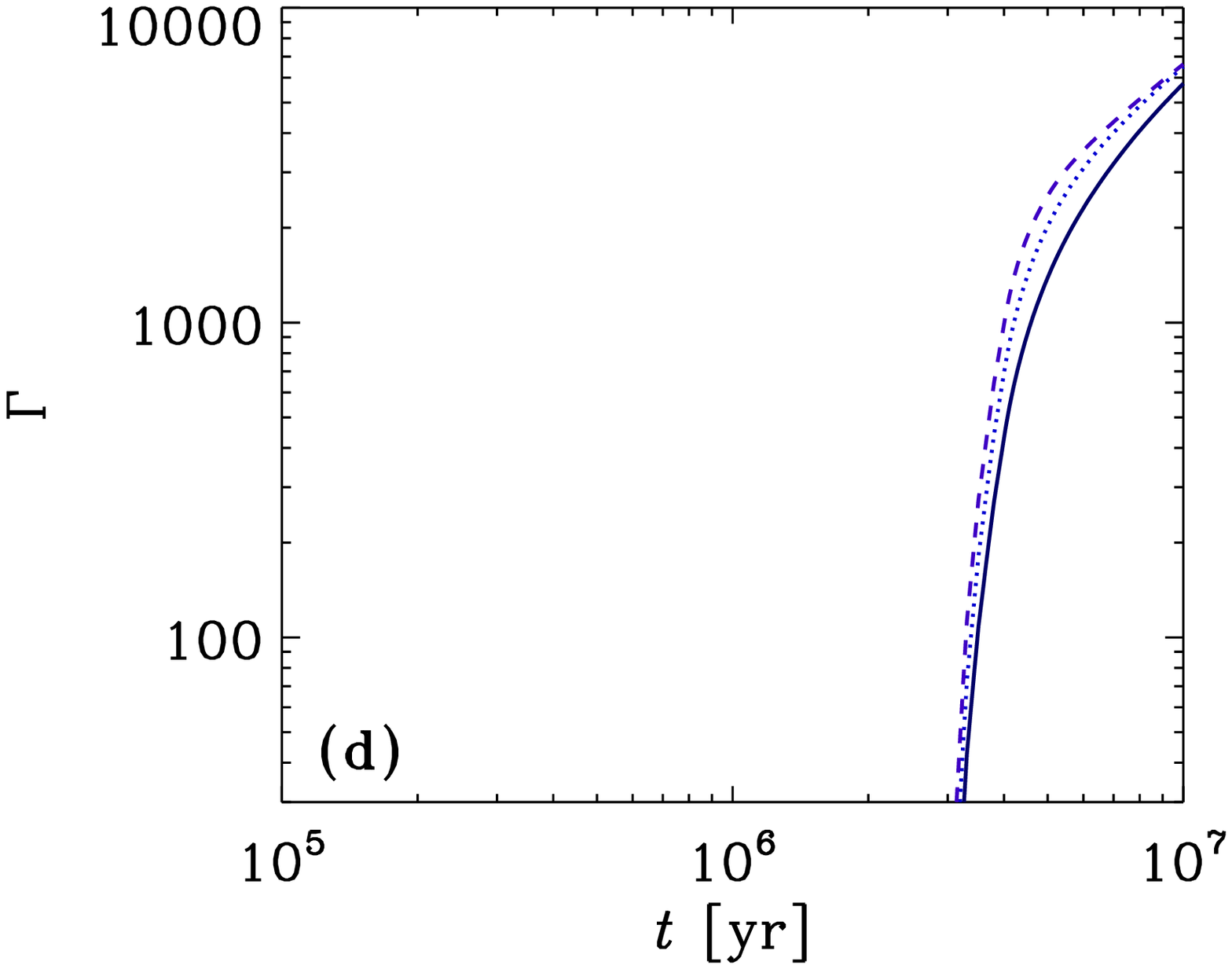}
\caption{Evolutions of the hydrogen number density
$n_\mathrm{H}$ (Panel a), the ionized radius
$r_\mathrm{i}$ (Panel b), the number of ionizing
photons emitted per unit time ${N}_\mathrm{ion}$
(Panel c), and
the cumulative number of SNe $\Gamma$
(Panel d). Solid,
dotted, and dashed lines are for
$n_\mathrm{H0}=3\times 10^4$, $10^5$,
and $3\times 10^5$ cm$^{-3}$, respectively, with
$M_0=3\times 10^6~\mathrm{M}_{\sun}$.
\label{fig:nH_ri}}
\end{figure*}

Submm emission in galaxies is usually dominated by
dust thermal radiation and
contaminated by free--free emission. In order to
estimate the contribution from free--free emission,
we use the VLA 2 cm continuum data, which is
sensitive to spatial scales up to 4 arcsec
\citep{beck02}. We adopt the emission on this scale
as the
contribution from the central star-forming region
{associated with the SSCs.}
At this
wavelength, the single-dish flux is 18.5 mJy, while
the VLA flux is $14\pm 1.5$ mJy
{(Table \ref{tab:data}; \citealt{beck02})}; thus,
76 per cent of the total radiation
comes from the central star-forming region at 2 cm.
Since the emission at 2 cm ($14\pm 1.5$ mJy) is dominated
by free--free emission \citep{beck02}, we extrapolate
the flux to
880 $\micron$
by assuming a frequency dependence of
$\propto\nu^{-0.1}$ \citep{osterbrock89}
and estimate the
contribution of free--free emission to the
880 $\micron$ flux to be
$10\pm 1$ mJy. Thus, $74^{+12}_{-10}$ per cent of the
flux detected at 880 $\micron$ by SMA is free-free
emission from the central star-forming region.
The contribution from free--free emission is
further investigated by
theoretical models in Section \ref{sec:model}.
The residual ($4\pm 2.5$ mJy) is possibly
dust emission or diffuse (more extended than $4''$ but
less than $\sim 10''$) free--free emission.


\section{Theoretical Models}\label{sec:model}

In order to {characterize} the radio continuum
radiation from a star-forming region, thermal
free--free radiation from \hii\ regions,
non-thermal synchrotron emission from supernova
remnants (SNRs), and dust thermal emission should
generally be taken into account. Here, we adopt
theoretical models for thermal free--free and
non-thermal synchrotron components from
\citet{hirashita06} to interpret the emission from
the central star-forming region in II~Zw~40.
Below we briefly summarize their models. We adopt
the same values for the physical parameters as those
in \citet{hirashita06} unless otherwise stated.
We also add a simple
calculation for dust emission in
Section \ref{sec:firsed}.

\subsection{Basic setups for the star formation}

The star formation is assumed to occur in a uniform
star-forming region, whose initial number density
of hydrogen nuclei is $n_\mathrm{H0}$. The mass
finally converted into stars, $M_0$, is also given.
The star formation rate (SFR) as a
function of time, $\psi (t)$, is written as
$\psi (t)=(\epsilon_\mathrm{SF}M_0/{\tff})
\, e^{-\epsilon_\mathrm{SF}t/\tff}$, where $t$ is
the time measured from the onset of the star
formation (thus, we assume $\psi =0$ for $t<0$),
and $\tff$ is the free-fall time-scale estimated by
$n_\mathrm{H0}$
(i.e.\ the SFR is regulated
by the dynamical time-scale; e.g.\
\citealt{elmegreen00}).
We simply introduce an efficiency
factor $\epsilon_\mathrm{SF}$, which
means that the gas is
converted into stars on a time-scale of
$\tff /\epsilon_\mathrm{SF}=1.4\times 10^6
(\epsilon_\mathrm{SF}/0.1)^{-1}
(n_\mathrm{H0}/10^5~\mathrm{cm}^{-3})^{-1/2}$ yr.
We treat $n_\mathrm{H0}$ and $M_0$ as independent
parameters. Throughout this paper, we assume a
Salpeter initial mass function (IMF) with a stellar
mass range of 0.1--100 $\mathrm{M}_{\sun}$.

{Since star formation does not proceed on a
single free-fall time-scale because of
some regulation factors such as
magnetic fields, turbulence, radiation, etc.\
\citep[e.g.][]{price09}, we tentatively
adopt $\epsilon_\mathrm{SF}=0.1$
\citep*[e.g.][]{lada10}}.\footnote{{The
efficiency factor $\epsilon_\mathrm{SF}$ is
different from the star formation efficiency,
which is usually defined as the fraction of
the finally formed stellar mass to the total
initial gas mass. In \citet{lada10}'s
notation, $\epsilon_\mathrm{SF}$ corresponds to
$1/f\sim 0.18$.}}
{Although $\epsilon_\mathrm{SF}$ may
be uncertain,
the quantities actually constrained
by the observational data are the gas mass
converted into stars and the gas density
(see Section \ref{subsec:result}). As long as
these quantities are focused on, the results
are insensitive to $\epsilon_\mathrm{SF}$.
Moreover, $\tff /\epsilon_\mathrm{SF}$ is
shorter than
the lifetimes of massive stars for most of
the parameter range in this paper; in this
case, the star formation is regarded as instantaneous
as long as we consider the evolution at
$t\ga\tff/\epsilon_\mathrm{SF}$. If the star
formation is instantaneous,
total stellar mass $M_0$, not the
detailed star formation history, determines
the total luminosity.
Special remarks will be made when we discuss ages
too young ($\la 1$ Myr) to regard the star formation as
instantaneous.}

\subsection{Radio SED}

Thermal free--free radiation is modeled by
considering the evolution of the size and density of
\hii\ region. In this paper, the recombination
coefficient excluding captures to
the ground level is assumed to be
$\alpha^{(2)}=1.73\times 10^{-13}~\mathrm{cm^3~s^{-1}}$
\citep[e.g.][]{spitzer78}.
We assume the density
in the \hii\ region to be uniform, and
neglect the effects of dust in the \hii\ region.
For details on the assumptions,
we refer to \citet{hirashita06}.

We consider the time evolution of the hydrogen number
density in the \hii\ region, $n_\mathrm{H}$, and
the radius of the ionized region, $r_\mathrm{i}$,
according to the evolution of the number of ionizing
photons emitted per unit time, $N_\mathrm{ion}$. We
consider the decrease of $n_\mathrm{H}$ by
pressure-driven expansion, which is assumed to
continue until $r_\mathrm{i}$ starts to decrease
by the death of massive stars. In calculating the SED
of thermal free--free component, the
gas temperature in the \hii\
region is assumed to be $T=13,000$ K \citep{thuan05}.
We also take free--free absorption into account.
In particular, for an \hii\ region in such a
dense region as
the II Zw 40 centre, free--free
absorption is important. The free--free optical depth
$\tau_\mathrm{ff}$ is
estimated by using the emission measure
EM as \citep{hunt04}
\begin{eqnarray}
\tau_{\rm ff} \simeq 0.328\left(
\frac{T}{10^4\,{\rm K}}\right)^{-1.35}\left(
\frac{{\rm EM}}{10^6\,{\rm pc\, cm}^{-6}}\right)
\left(\frac{\nu}{\rm GHz}\right)^{-2.1},
\label{eq:tau_ff}
\end{eqnarray}
where EM is
estimated in terms of the electron
number density $n_e$ and the radius of ionized
region $r_\mathrm{i}$ as
$\mathrm{EM}=4n_e^2r_\mathrm{i}/3$.
We assume that the absorbing medium is intermixed
with the emitting material; that is,
the escape fraction is assumed to be
$[1-\exp (-\tau_{\rm ff})]/{\tau_{\rm ff}}$.


The non-thermal radio luminosity of SNRs is estimated
as
$L_\mathrm{nt}^0(\nu )=l_\mathrm{nt}\tau_\mathrm{nt}
(\nu /5~\mathrm{GHz})^{-0.5}\gamma (t)$,
where $\gamma (t)$ is the rate of core-collapse
supernovae (SNe)\footnote{In this paper, we only
consider core-collapse SNe originating from massive
stars.} estimated by assuming that stars more massive
than 8 M$_{\sun}$ become SNe,
and $l_\mathrm{nt}\tau_\mathrm{nt}$ is the radio
energy emitted by a SNR over its entire lifetime
(fluence) at 5 GHz. The fluence possibly depends
on the ambient density
\citep{arbutina05}. We adopt
$l_\mathrm{nt}\tau_\mathrm{nt}(5~\mathrm{GHz})=
7.6\times 10^{22}$ W Hz$^{-1}$ yr, since this value
fits the radio SED of SBS 0335$-$052, which also
hosts dense and compact \hii\ regions
\citep{hirashita06}.
Free--free absorption is also applied for all
the non-thermal component by a screen geometry:
i.e.\ the observed non-thermal luminosity is
$L_\mathrm{nt}=L_\mathrm{nt}^0\,
e^{-\tau_\mathrm{ff}}$.
This minimizes the contribution from the
non-thermal component especially at low
frequencies. If the non-thermal component
is not absorbed efficiently, the existence of
the non-thermal component is more severely
excluded for the II~Zw~40 centre
(see Section \ref{subsec:result}).
{Synchrotron self-absorption is not
important for SNRs in the frequency
range of interest \citep[e.g.][]{condon92}.}

We also define
the cumulative number of SNe, $\Gamma$, as
\begin{eqnarray}
\Gamma (t)=\int_0^t\gamma (t')\,dt'.
\end{eqnarray}
$\Gamma (t)$ is used as an indicator of the
dust production in SNe
(Section \ref{subsec:origin}).

\subsection{Theoretical results for the radio SED}
\label{subsec:result}

We present the results calculated by the framework
described above. We concentrate on the parameter
ranges relevant for the central star formation
in II Zw 40. As shown later,
$n_\mathrm{H0}\sim 10^5$\,cm$^{-3}$ and
$M_0\sim 3\times 10^6$\,M$_{\sun}$
fit the radio SED if such a young age $\sim 3$ Myr
as indicated by
the optical observations (see Introduction) is
adopted. The radio luminosity is almost proportional
to $M_0$, so the constraint on $M_0$ is
rather severe. Thus, we mainly investigate the case of
$M_0=3\times 10^6~\mathrm{M}_{\sun}$. For the density,
we examine $n_\mathrm{H0}=3\times 10^4$, $10^5$,
and $3\times 10^5$ cm$^{-3}$.

In Fig.\ \ref{fig:nH_ri}, we show the key quantities
($n_\mathrm{H}$, $r_\mathrm{i}$,
${N}_\mathrm{ion}$, and $\Gamma$) as functions of
time. The ionized region radius $r_{\rm i}$
monotonically increases and the density
$n_\mathrm{H}$ decreases until $t\sim 3$ Myr because
of the pressure-driven
expansion and the increase of ${N}_{\rm ion}$.
After that, the expansion stops
because $N_\mathrm{ion}$ decreases. The initial SFR is
$\psi (0)=\epsilon_\mathrm{SF}M_0/
t_\mathrm{ff}\simeq 2.2(\epsilon_\mathrm{SF}/0.1)
(M_0/3\times 10^6~\mathrm{M}_{\sun})
(n_\mathrm{H0}/3\times 10^5~\mathrm{cm}^{-3})^{1/2}~
\mathrm{M}_{\sun}~\mathrm{yr}^{-1}$.
The SFR measured by the H$\alpha$ luminosity is roughly
$\sim 1~\mathrm{M}_{\sun}$ yr$^{-1}$
\citep{vanzee98,vanzi08}.
Since the SFR exponentially decays on a time-scale of
$\tff/\epsilon_\mathrm{SF}\sim 1.4
(\epsilon_\mathrm{SF}/0.1)^{-1}
(n_\mathrm{H0}/10^5~\mathrm{cm}^{-3})^{-1/2}$ Myr
in our models, we
obtain the SFR averaged for 3 Myr as
$\sim 0.91~\mathrm{M}_{\sun}$ yr$^{-1}$
($n_\mathrm{H0}=10^5$ cm$^{-3}$ is assumed),
which is near to the value obtained from the H$\alpha$
line. As shown later, the SFR assumed also reproduces
the interferometric radio continuum flux.
For the stellar mass,
\citet{buckalew05} derived
$6.3\times 10^{6}~\mathrm{M}_{\sun}$ from a
stellar spectral synthesis model
\citep[Starburst 99;][]{leitherer99} with a Salpeter
IMF of stellar mass range 1--120 $\mathrm{M}_{\sun}$ and a
metallicity of 1/5 solar.
\citet{vanzi08} derived $1.7\times 10^6~\mathrm{M}_{\sun}$ with
the same spectral synthesis model but with a Kroupa IMF.
Our stellar mass ($M_0\sim 3\times 10^6~\mathrm{M}_{\sun}$) is
bracketed by those two results,
which means that our stellar mass is consistent with
those in the literature within the uncertainty in the
IMF. The evolution of $\Gamma$ is used for the
discussion of dust production by SNe in
Section~\ref{subsec:origin}. $\Gamma$ rapidly increases
after $t=3$ Myr, when the first SNe occur.

\begin{figure}
\includegraphics[width=0.48\textwidth]{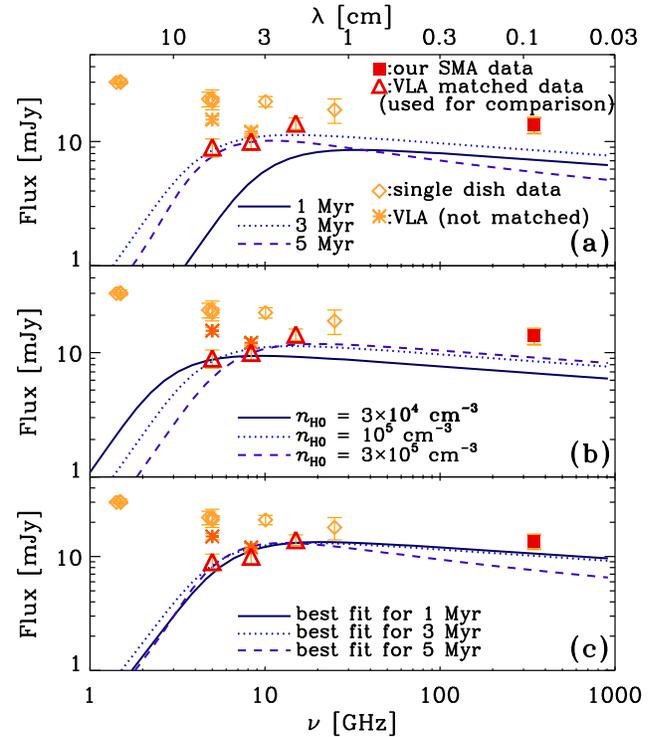}
\caption{Radio SEDs. (a) Solid, dotted, and dashed
lines show the SEDs at 1, 3, and 5 Myr, respectively.
The initial hydrogen number density and the total
mass converted into stars are fixed with
$n_\mathrm{H0}=10^5$ cm$^{-3}$ and
$M_0=3\times 10^6~\mathrm{M}_{\sun}$,
respectively. The observational data
(Table \ref{tab:data}) are also presented. The filled
square represents our SMA
measurement at $880~\micron$ for the central
star-forming region. Diamonds show single-dish
measurements by \citet{jaffe78}, \citet{klein84},
\citet{klein91}, while asterisks and triangles
indicate the VLA interferometry data by
\citet{beck02}. The triangles at 6, 3.6, and 2 cm are
the matched VLA fluxes obtained by restricting
$(u,\, v)$ data to baselines greater than 20k$\lambda$
(i.e.\ sensitive to the structures smaller than
4 arcsec) \citep{beck02}. The asterisks at 6 and 3.6 cm
are the VLA fluxes synthesized by all the $(u,\, v)$
data, which are sensitive to structures up to 10 arcsec and
7 arcsec, respectively. (b) Same as Panel (a) but for
the SEDs at
3~Myr with various initial densities. Solid, dotted,
and dashed lines are for $n_\mathrm{H0}=3\times 10^4$,
$10^5$, and $3\times 10^5$ cm$^{-3}$, respectively
($n_\mathrm{H}=1.8\times 10^3$, $3.2\times 10^3$, and
$5.4\times 10^3$\,cm$^{-3}$ at 3 Myr, respectively).
(c) Same as Panel (a) but for the
best-fit model parameters ($M_0$ and $n_\mathrm{H0}$)
listed in
Table \ref{tab:fit}. The fitting is applied for the
three triangles (the matched VLA fluxes). The solid,
dotted,
and dashed lines show the best-fit solutions for
$t=1$, 3, and 5 Myr, respectively.
\label{fig:sed_age}}
\end{figure}

In Figure \ref{fig:sed_age}a, we show the radio SEDs
at $t=1$, 3, and 5\,Myr with
$n_\mathrm{H0}=10^5$ cm$^{-3}$ and
$M_0=3\times 10^6~\mathrm{M}_{\sun}$. As the age
becomes older, the peak shifts to lower frequencies,
because the free-free optical depth becomes smaller
as the \hii\ region expands. At $t=1$ and 3 Myr, the
emission is completely dominated by free--free
emission, and at $t=5~\mathrm{Myr}$, the
synchrotron component begins to contribute to the
emission and the spectrum slope changes.
For comparison, the VLA `matched' data whose
$(u,\, v)$ coverage is restricted to baselines
greater than 20k$\lambda$ (i.e.\ sensitive to
structures smaller than 4 arcsec) are adopted
(three triangles in Fig.\ \ref{fig:sed_age})
as representative fluxes from the central
star-forming region.

The density strongly affects the frequency at which
the flux peaks because free--free absorption is
sensitive to the density. In Fig.\ \ref{fig:sed_age}b,
we show the SEDs {at $t=3$ Myr} for various
initial densities ($n_\mathrm{H0}$) with
$M_0=3\times 10^6~\mathrm{M}_{\sun}$. We observe that
the peak position of the SED is indeed
sensitive to the density. The rising
spectrum of the matched data (triangles) is
consistent with free--free absorption.

It is possible to search for the best-fit values of
$M_0$ and $n_\mathrm{H0}$ for each age. The
matched VLA data are adopted (three triangles in
Fig.\ \ref{fig:sed_age}) for the $\chi^2$ fitting (one
degree of freedom), since our models are applicable
to the central
star-forming region. The best-fit solutions are shown
in Table \ref{tab:fit} and Fig.\ \ref{fig:sed_age}c.
In the case of $t=1$ Myr, a large $M_0$ is required
since only 29 per cent of $M_0$ is converted into
stars at $t=1$~Myr.
As seen in Fig.\ \ref{fig:sed_age}c, the
spectral slope at $\nu\ga 15~\mathrm{GHz}$ for
$t=5$ Myr is different from that at
$t\leq 3$ Myr because of the
contribution from the non-thermal component.
However, the spectral slope at $\nu\ga 15$ GHz
in the central part of II Zw 40 is not well
constrained because of the lack of the high-resolution
data at millimetre wavelengths.
For all the three ages, a stellar mass of
$\sim 3$--$4\times 10^6~
\mathrm{M}_{\sun}$\footnote{At $t=1$ Myr,
only 29 per cent of $M_0$ (i.e.\
$3.2\times 10^6$ M$_{\sun}$) is converted into stars,
while at $t\ga 3$ Myr, most of $M_0$ is converted
into stars (Table \ref{tab:fit}).} and a current
number density in the \hii\ region of
$4$--$5\times 10^3$ cm$^{-3}$ are obtained for
the best-fit values.

\begin{table}
\centering
\begin{minipage}{70mm}
\caption{Best-fit solutions.}
\label{tab:fit}
\begin{tabular}{ccccc}
\hline
Age & $M_0$ & $n_\mathrm{H0}$ & $n_\mathrm{H}$  &
$\chi^2$ \\
(Myr) & ($\mathrm{M}_{\sun}$) & (cm$^{-3}$)
& (cm$^{-3}$) & \\
\hline
1 & $1.1\times 10^7\,^\mathrm{a}$ & $2.2\times 10^4$
& $4.7\times 10^3$ & 2.6 \\
3 & $3.5\times 10^6$ & $1.2\times 10^5$ & $3.7\times 10^3$
& 2.0 \\
5 & $4.1\times 10^6$ & $1.2\times 10^5$ & $3.9\times 10^3$
& 2.9 \\
\hline
\end{tabular}

\medskip

$^\mathrm{a}$
{Only 29\% of $M_0$ (i.e.\ $3.2\times 10^6$ M$_{\sun}$)
is converted into stars at $t=1$ Myr.}
\end{minipage}
\end{table}

If the age is about 3 Myr as suggested from the
optical and near-infrared observations
(Introduction; \citealt{vanzi08}), 75 per cent of the
$880~\micron$ flux obtained in our observation is
explained by free--free emission according to the
best-fit SED at 3 Myr (the flux at $880~\micron$ is
10.2\,mJy in the model, while the observed flux is
13.6\,mJy). The difference ($\simeq 3.4$ mJy) is
likely to be due to the dust and/or diffuse (i.e.\ not
associated with the
compact \hii\ region) free--free emission.
The dust emission in the
central star-forming region is
modeled and examined in the next section.

\section{Contribution from dust in
the centre}\label{sec:firsed}

\subsection{Dust emission associated with the
central star-forming region}\label{subsec:dust}

Here we model the dust emission associated with
the central star-forming region in II Zw 40. We do
not model
the entire system, so that the following FIR
luminosity and dust temperature are valid for
the central part of II Zw 40, and are not
representative for the global quantities.
We consider the wavelength range where
large grains, which achieve radiative
equilibrium with the ambient stellar radiation field,
govern the emission
($\lambda\ga 100~\micron$), since
the large grains dominate the total dust
mass \citep[e.g.][]{galliano05}.
We do not treat very small grains
contributing to mid-infrared emission.

For simplicity, we adopt a shell-like geometry for
the dust distribution. 
This simplification, which is
also assumed in \citet{takeuchi03,takeuchi05} and
\citet{galliano05},
helps to decrease the number of physical parameters
that affect the emergent dust emission SED in
relatively minor ways. If the dust is distributed in
a thin shell
at a distance $R_\mathrm{dust}$ from the centre,
the dust optical depth, $\tau_\mathrm{dust}$, for the
radiation from stars is estimated as
\begin{eqnarray}
\tau_\mathrm{dust}=
\frac{3M_\mathrm{dust}}{16\pi R_\mathrm{dust}^2as},
\label{eq:taudust}
\end{eqnarray}
where $M_\mathrm{dust}$ is the total dust mass
in the shell, $a$ is the grain radius (uniform
spherical grains are assumed), and $s$ is the
grain material density. Note that $R_\mathrm{dust}$
will typically be larger than
$r_\mathrm{i}$ since the dust associated with
surrounding neutral or molecular gas contributes
significantly to the FIR luminosity \citep{xu92}.
In this paper, we adopt
$a=0.1~\micron$ and $s=3$ g cm$^{-3}$
{\citep[e.g.][]{draine84}}. We estimated
the absorption cross section of a grain with the
geometrical one because the major heating source
is ultraviolet (UV) {radiation} from OB stars
\citep[e.g.][]{buat96} {(i.e.\ the grain radius is
comparable to the wavelength)}.

We assume that the UV luminosity is equal to the
bolometric luminosity of the OB stars. The OB star
luminosity is estimated as a function of time:
\begin{eqnarray}
\mathcal{L}_\mathrm{OB}(t)=
\int_{3~\mathrm{M}_{\sun}}^\infty\mathrm{d}m
\int_0^{\tau_m}\mathrm{d}t'\, L(m)\,\phi
(m)\,\psi (t-t'),\label{eq:LOB}
\end{eqnarray}
where $L(m)$ is the main sequence luminosity of a star
with mass $m$ and $\phi (m)$ is the IMF.
For $L(m)$, we
adopt the zero metallicity case in \citet{schaerer02}
to consider a low-metallicity evolutionary stage,
and $L(m)$ may be 2 times smaller if we assume solar
metallicity.
We assume that the OB stars are
located at the
centre. This assumption is valid when we consider the
central SSCs as the source of UV radiation in II Zw 40.
Since the absorbed UV light is reprocessed into
FIR wavelengths, the dust FIR luminosity
is estimated by
\begin{eqnarray}
\mathcal{L}_\mathrm{FIR}=(1-e^{-\tau_\mathrm{dust}})
\mathcal{L}_\mathrm{OB}.\label{eq:radtr}
\end{eqnarray}
For simplicity, we adopt a single-temperature
approximation for the dust emission, so that the
monochromatic luminosity is expressed as
\begin{eqnarray}
L_\mathrm{FIR}(\nu )=4\pi\kappa_\nu M_\mathrm{dust}B_\nu
(T_\mathrm{dust}),\label{eq:dustsed}
\end{eqnarray}
where $B_\nu$ is the Planck function, $\kappa_\nu$ is
the mass absorption coefficient
of the dust, and $T_\mathrm{dust}$ is the dust
temperature. We assume that
$\kappa_\nu =0.7(\nu /340~\mathrm{GHz})^2$
cm$^2$ g$^{-1}$ \citep{james02}. The dust temperature is
determined so
that it satisfies the total luminosity constraint:
\begin{eqnarray}
\mathcal{L}_\mathrm{FIR}=\int_0^\infty L_\mathrm{FIR}
(\nu )\,\mathrm{d}\nu .\label{eq:dustsedinteg}
\end{eqnarray}
By combining
Equations (\ref{eq:radtr})--(\ref{eq:dustsedinteg}),
we obtain
\begin{eqnarray}
\left( 1-e^{-\tau_\mathrm{dust}}\right)
\mathcal{L}_\mathrm{OB}/M_\mathrm{dust}=1.09\times
10^{-5}T_\mathrm{dust}^6.\label{eq:dust_temp}
\end{eqnarray}

The dust emission SED, $L_\mathrm{FIR}(\nu )$, is
determined as follows. $R_\mathrm{dust}$ is given.
Then, a certain test value of $M_\mathrm{dust}$ is
assumed. By using Equation (\ref{eq:taudust}),
$\tau_\mathrm{dust}$ is obtained. Consequently
$T_\mathrm{dust}$ is derived from
Equation (\ref{eq:dust_temp}).
Note that the age has to be assumed to evaluate
$\mathcal{L}_\mathrm{OB}$ (Equation \ref{eq:LOB}).
Finally, the dust SED $L_\mathrm{FIR}(\nu )$ is
obtained by Equation (\ref{eq:dustsed}). If the
predicted flux
$L_\mathrm{FIR}(\nu )/(4\pi D^2)$ overproduces
(underproduces)
the observed flux, we decrease (increase)
$M_\mathrm{dust}$ and repeat the above procedure.



\subsection{Comparison with observational data}
\label{subsec:comparison}

\begin{figure}
\includegraphics[width=0.45\textwidth]{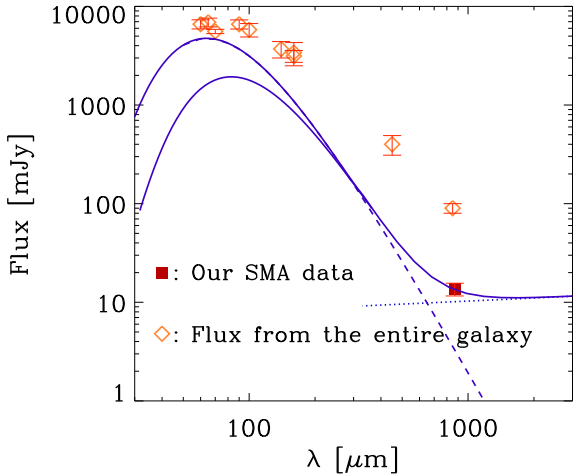}
\caption{FIR--submm SEDs. Solid, dotted,
and dashed lines represent, respectively, the total,
the free--free component, and the dust component
calculated by the models for the central region
in II Zw 40. For the free--free component,
the best-fit SED to the matched VLA data for
$t=3~\mathrm{Myr}$ (i.e.\ the dotted line in
Fig.\ \ref{fig:sed_age}c) is adopted. The dust mass
is adjusted to reproduce the SMA data point at
880 $\micron$ (filled square). Two cases for the
radius of dust distribution are examined:
$R_\mathrm{dust}=100$ and 250 pc for the upper and
lower solid lines, respectively. As upper limits for the
flux in the central region, we also show
observational data for the entire II Zw 40
system (open diamonds), which
are taken from
\citet{vader93} for 60 and 100 $\micron$ (\textit{IRAS}),
\citet{hunt05} for 450, and 850 $\micron$ (SCUBA),
\citet{galliano05} for 450 and 850 $\micron$
(SCUBA), \citet{engelbracht08} for 70 and
160 $\micron$ (\textit{Spitzer}),
\citet{hirashita08} for 65, 90, 140, and 160 $\micron$
(\textit{AKARI}). {The observational data
are summarized in Table \ref{tab:fir}.}
\label{fig:sed_fir}}
\end{figure}

As mentioned in Section \ref{subsec:result}, the
difference between the predicted flux and the measured
one by SMA at 880 $\micron$ should be dust
emission or diffuse free--free emission. To simplify
the discussion, we interpret that all the difference
is dust emission, which means that the following
values for $M_\mathrm{dust}$ and $L_\mathrm{FIR}$
should be taken as upper limits. In the SMA image
(Fig.\ \ref{fig:image}), the extension of the
880 $\micron$ emission is $\sim 10$ arcsec. That is,
$R_\mathrm{dust}\la 250$ pc.
Any more extended dust component contributes to the
difference between the SMA flux and the SCUBA flux.
Also, $R_\mathrm{dust}\la 60~\mathrm{pc}$ is
rejected because the
dust temperature is so high that the 60 $\micron$
flux exceeds the \textit{IRAS} flux. Thus, we
investigate $R_\mathrm{dust}=100$ and 250 pc
as representative cases.
For the free--free component, we adopt the best-fit
model for $t=3~\mathrm{Myr}$ (Table \ref{tab:fit}), i.e.\
$n_\mathrm{H0}=1.2\times 10^5~\mathrm{cm}^{-3}$
and $M_0=3.5\times 10^6~\mathrm{M}_{\sun}$
(the dotted line in Figure \ref{fig:sed_age}c).
For the dust emission, we adopt the dust SED model in
Section \ref{subsec:dust}, and we adjust
$M_\mathrm{dust}$ so that it reproduces the SMA
observational data according to the formulation
in Section \ref{subsec:dust}: we obtain
$(M_\mathrm{dust},\, T_\mathrm{dust})
=(1.8\times 10^4~\mathrm{M}_{\sun},\, 45~\mathrm{K})$ and
$(2.7\times 10^4~\mathrm{M}_{\sun},\, 35~\mathrm{K})$,
for $R_\mathrm{dust}=100$ and 250 pc, respectively.
We present the SEDs in Figure \ref{fig:sed_fir}.
As upper limits for the flux from the central part,
we also show the data taken by
\textit{IRAS} \citep{vader93},
\textit{AKARI} \citep{hirashita08}, \textit{Spitzer}
\citep{engelbracht08} and SCUBA
\citep*{galliano05,hunt05} for the fluxes
from the entire system.
Although mid-infrared
\textit{Spitzer} data are available with resolutions
comparable to our SMA data, mid-infrared emission
comes from a different dust component,
very small grains, as mentioned in
Section \ref{subsec:dust}.
Thus, we only concentrate on FIR and submm
wavelengths where large grains
dominate the dust emission.

Fig.\ \ref{fig:sed_fir} shows that a significant fraction
of the 60--100 $\micron$
flux (71 per cent of the \textit{IRAS} 60 $\micron$
flux and 54 per cent of the \textit{IRAS} 100 $\micron$
flux) is explained by the emission from the central
star-forming region for
$R_\mathrm{dust}=100$ pc.
If we adopt $R_\mathrm{dust}=250$ pc,
the contribution from the central star-forming
region is smaller (22 per cent of the
\textit{IRAS} 60 $\micron$ flux and 31 per cent
of the \textit{IRAS} 100 $\micron$ flux)
because of a lower dust
temperature.
In any case, an
additional emission from a diffuse dust component is
necessary to explain the FIR fluxes
of the entire system, {especially,}
at wavelengths
longer than 100 $\micron$.


\subsection{The origin of dust in the centre}
\label{subsec:origin}

Now we consider the origin of dust in the II Zw 40
centre. One of the natural explanations for the
dust mass derived above
($\sim 2$--$3\times 10^4\,\mathrm{M}_{\sun}$)
is production in previous episodes of
star formation (i.e.\ preexisting). First, before
investigating the possibility of
preexisting dust, we examine how much dust
can be produced in the current star formation activity.
Because the age is young, SNe are
the only source of dust grains and evolved low-mass
stars such as
asymptotic giant branch (AGB) stars
are negligible
\citep{valiante09,gall11}.
Theoretical studies
suggest that $\sim 0.1$--1 $\mathrm{M}_{\sun}$ of dust grains
condense in a SN
\citep{kozasa89,todini01,nozawa03,bianchi07,nozawa07,nozawa10}.
Infrared and submm observations of SNRs
also detected
$\sim 0.01$--1 $\mathrm{M}_{\sun}$ of dust
\citep[e.g.,][]{rho09,dunne09,gomez09,matsuura11}, although
quantitative significance of SNe to the dust
enrichment in the early galaxy evolution is still
debated \citep[][and references therein]{nozawa10}.
If we adopt
1 $\mathrm{M}_{\sun}$ as the `maximum' dust yield
per SN, the total dust mass
1--$3\times 10^4$ M$_{\sun}$ is explained only if
at least $10^4$ SNe occur.
(If we consider dust destruction in SNe,
more SNe are
necessary to explain the total dust amount.)
According to
Fig.\ \ref{fig:nH_ri}, $\Gamma$ does
not reach this number even at $t=10^7$ yr.
Thus, we conclude that the excess at
880 $\micron$ observed in
the centre of II Zw 40, if the emission comes
from dust, should be contributed from preexisting
dust.

As for the possibility of preexisting dust,
we may be observing the preexisting
dust itself, or
dust grains after the growth by the accretion of
metals.
\citet{hirashita11} derive the dust growth
time-scale for silicate (a similar time-scale is
obtained for carbonaceous dust):
\begin{eqnarray}
\tau_\mathrm{grow} & \simeq & 2.1\times 10^7~
\mathrm{yr}\left(
\frac{\langle a^3\rangle /\langle a^2\rangle}{0.1~\micron}
\right)
\left(\frac{Z}{1~\mathrm{Z}_{\sun}}\right)^{-1}\nonumber\\
& &
\times\left(\frac{n_\mathrm{H}}{10^3~\mathrm{cm}^{-3}
}\right)^{-1}
\left(\frac{T_\mathrm{gas}}{50~\mathrm{K}}\right)^{-1/2}
\left(\frac{S}{0.3}\right)^{-1}\, ,
\end{eqnarray}
where $\langle a^3\rangle$ and $\langle a^2\rangle$
are the averages of $a^3$ and $a^2$ ($a$ is the
grain radius) for grain size distribution, $Z$ is
the metallicity, $n_\mathrm{H}$ is the
hydrogen number density, $T_\mathrm{gas}$ is the gas
temperature, and $S$ is the sticking efficiency of
the relevant metal species onto the dust surface.
We assume
$\langle a^3\rangle /\langle a^2\rangle =0.1~\micron$,
$Z=1/4$ Z$_{\sun}$ (see Introduction),
$n_\mathrm{H}=10^5~\mathrm{cm}^{-3}$
(Table \ref{tab:fit}), $T_\mathrm{gas}=50$ K
{\citep{wilson97}},
and $S=0.3$ {\citep{leitch85,grassi11}}.
Then, we obtain
$\tau_\mathrm{grow}\sim 8.4\times 10^5$ yr,
which is comparable to the star formation
time-scale. Therefore,
if the preexisting grains survive before the
current star formation
episode, they can grow by the accretion of
gas-phase metals.



\section{Radio--FIR relation}\label{sec:discuss}

The 880 $\micron$ emission in the II Zw 40 centre
is dominated by free-free emission. At this
wavelength, the emission is usually dominated
by dust on a galactic scale.
{Thus, we expect that the properties of
FIR dust emission and
radio free--free emission in the II Zw 40 centre
are different from global galaxy
properties. In this section, we examine this
expectation quantitatively by using the results
in the above sections.}
Hereafter, we use the
term ``FIR luminosity'' for the dust emission luminosity
integrated over {8--1000 $\micron$}.

\begin{figure}
\includegraphics[width=0.45\textwidth]{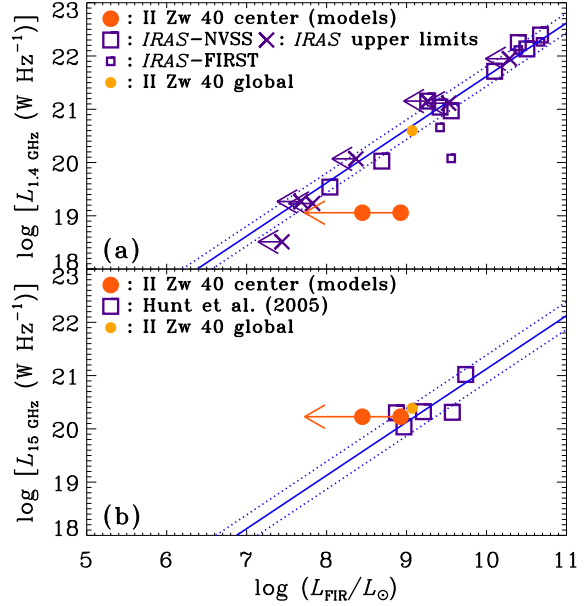}
\caption{Radio--FIR relation for the central part of
II Zw 40 (large filled circles) in terms of the
global relations for BCDs. Radio frequencies of 1.4
and 15 GHz are adopted in Panels (a) and (b),
respectively. Two cases for the radius of dust
distribution are presented for $L_\mathrm{FIR}$:
$R_\mathrm{dust}=100$ and 250 pc for the right and
left points (corresponding the two SEDs in
Figure \ref{fig:sed_fir}). The arrow
indicates that the FIR luminosities
estimated from the models are taken as upper limits
since all the excess over the model free--free
emission is assumed to be dust emission.
The observational data for the global emission from
BCDs in Panel (a) are taken from \citet{hopkins02}
(large/small open squares are for NVSS/FIRST data
and crosses are for non-detection at 100 $\micron$),
while those in Panel (b) are taken from \citet{hunt05}.
The entire II Zw 40 system is also plotted
(1.4 GHz, 15 GHz, and FIR data are taken from
\citealt{jaffe78}, \citealt{klein91}, and
\citealt{vader93}, respectively).
We also show the lines with $q_{1.4}=2.39\pm 0.19$
in Panel (a) and $q_{15}=2.89\pm 0.26$ in Panel (b)
(solid lines with dotted lines for 1 $\sigma$)
to indicate how much the central part of II Zw 40 is
deviated from the reference radio--FIR relation defined
for the BCDs.
\label{fig:radio_fir}}
\end{figure}

In Fig.\ \ref{fig:radio_fir}, we show the relation
between the monochromatic luminosity at 1.4 GHz
($L_\mathrm{1.4\,GHz}$) and the FIR luminosity
$L_\mathrm{FIR}$ for the central star-forming region
in II Zw 40. The radio luminosity at 1.4 GHz is adopted
from the best-fit model for $t=3~\mathrm{Myr}$
(Figure \ref{fig:sed_age}c), while the FIR
luminosity is evaluated for $R_\mathrm{dust}=100$ and
250 pc (Section \ref{subsec:comparison};
Fig.\ \ref{fig:sed_fir}).
Since all the difference between the model free--free
flux and the observed 880 $\micron$ flux is assumed to
come from dust, the FIR luminosities estimated from
the models should be taken as upper limits.

For comparison, we also plot the observational data
of BCDs in Fig.\ \ref{fig:radio_fir}. The sample is
taken from \citet{hopkins02} for the FIR and 1.4 GHz
luminosities of the entire
system (i.e.\ global luminosities).
We only adopt the
objects with detections by the NRAO VLA Sky Survey
(NVSS) at $\nu =1.4$ GHz \citep{condon98}
and by the \textit{IRAS} at $\lambda =60~\mu$m
(large open squares).
We also adopt the radio flux
measured by the Faint Images of the Radio Sky at
Twenty cm \citep[FIRST;][]{becker95}
(small open squares).
The FIRST flux is systematically smaller
than the NVSS flux. The probable reason which
\citet{hopkins02} suggest is the
difference in their sensitivity to extended emission:
FIRST images tend to miss extended emission whose
angular size is larger than about 2 arcmin.
The FIR luminosity of the BCD sample is
estimated based on the \textit{IRAS} 60 $\mu$m and
100 $\mu$m fluxes obtained from NASA/IPAC
Extragalactic Database (NED).
{As an observational estimate of the
FIR luminosity $L_{\rm FIR}$, we adopt
an empirically
derived formula by \citet{nagata02}, who estimate
the total dust
luminosity at $\lambda\geq 40~\micron$
by using the \textit{IRAS} 60 and 100 $\micron$
fluxes. \citet{nagata02} take into account
the contribution from wavelengths longer than the
\textit{IRAS} bands  by integrating
the modified blackbody spectrum, but they do not
consider the contribution from $\lambda <40~\micron$.
Therefore, the FIR fluxes may be underestimated by
$<30$ per cent. However,
\citet{nagata02}'s method can avoid a significant
underestimate in a simple \textit{IRAS} luminosity
by \citet{helou88}, who only considered the contribution
from the luminosity in the \textit{IRAS} FIR
bands.\footnote{For II Zw 40, the global FIR luminosity
estimated by \citet{nagata02}'s method is
$1.2\times 10^9$ L$_{\sun}$,
while that estimated by \citet{helou88}'s method is
$9.9\times 10^8$ L$_{\sun}$.}} If a sample BCD is not
detected at
$100~\micron$, we utilize the upper limit at
$100~\micron$ to estimate an upper limit of
$L_{\rm FIR}$. The data with upper limits are shown
by crosses in Figure \ref{fig:radio_fir}
($L_{\rm 1.4~GHz}$ is taken from the NVSS data).
We also show the global luminosities of II Zw 40, whose
1.4 GHz and FIR data are taken from \citet{jaffe78}
and \citet{vader93}, respectively.

In order to quantitatively discuss the possible
deviation from the standard radio--FIR relation
for BCDs, we adopt the radio-to-FIR ratio
as usually used \citep[e.g.][]{condon92}. Here
we define $q_{1.4}$ and $q_{15}$ as
\begin{eqnarray}
q_{1.4} & \equiv & \log\left(
\frac{L_\mathrm{FIR}}{3.75\times 10^{12}~\mathrm{W}}
\right)
-\log\left(
\frac{L_{1.4~\mathrm{GHz}}}{\mathrm{W~Hz}^{-1}}
\right) ,\\
q_{15} & \equiv & \log\left(
\frac{L_\mathrm{FIR}}{3.75\times 10^{12}~\mathrm{W}}
\right)
-\log\left(
\frac{L_{15~\mathrm{GHz}}}{\mathrm{W~Hz}^{-1}}
\right) .
\end{eqnarray}
The average and the standard deviation of
$q_{1.4}$ are calculated for the sample detected
by both \textit{IRAS} (60 and 100 $\micron$) and
NVSS, while those of $q_{15}$ are determined for
the sample in \citet{hunt05}. The averages of
$q_{1.4}$ and $q_{15}$ are
2.39 and 2.89, respectively,
and the standard deviations are 0.19 and 0.26,
respectively. We also show the lines with
$q_{1.4}=2.39\pm 0.19$ and $q_{15}=2.89\pm 0.26$
to show how much the radio--FIR relation in the
central part of II Zw 40 is
deviated from the reference radio--FIR relation
for the BCDs.

{}From Fig.\ \ref{fig:radio_fir}, we observe that
the central region of II Zw 40 has a suppressed
1.4 GHz luminosity and is located below the
radio--FIR relation of global BCDs, if the FIR
luminosity of
II Zw 40 has a value near to the upper limit.
This is interpreted to
be due to free--free absorption in our models
(Fig.\ \ref{fig:sed_age}).
{Our SMA observation is
not sensitive to the possible diffuse synchrotron
radiation that would not be affected by free--free
absorption. Indeed, the relatively high 1.4 GHz
flux from the entire galaxy (Fig.\ \ref{fig:sed_age})
indicates the presence of diffuse nonthermal emission
in this galaxy. The trend that compact sources have
lower radio-to-FIR ratios
is also found for luminous infrared galaxies by
\citet{condon91}, who also attribute the lower ratios to
free--free absorption.}

To avoid free--free absorption, we use the 15 GHz
luminosity. The best-fit model for $t=3$ Myr is used
for the 15 GHz luminosity of the II Zw 40 centre
(Section \ref{subsec:comparison}). This luminosity
fits the observational data at 15 GHz by
\citet{beck02}. For comparison, we show the
global luminosities of BCDs
compiled by \citet{hunt05}
{($L_\mathrm{FIR}$ is estimated
by the method of \citealt{nagata02}; see above)}.
{As we observe in Fig.\ \ref{fig:radio_fir}b,
there is a possibility that
the radio--FIR relation is consistent with
the global relation if the dust distribution is
as compact as $R_\mathrm{dust}=100$ pc
(but note that the FIR luminosity is upper limit).}
If the dust distribution is as extended
as $R_\mathrm{dust}=250~\mathrm{pc}$, the FIR luminosity is
significantly suppressed (or the radio luminosity is
enhanced) in the centre of II Zw 40.

The deviation of the II Zw 40 centre in the
radio--FIR relation is actually expected, rather
than peculiar.
Indeed, \citet{cannon05} find a spatial variation
of FIR-to-radio ratio by an order of
magnitude within a metal poor dwarf
galaxy, IC 2574. \citet{dumas11} also
show different radio--FIR relations between
spiral arms and interarm regions in M51.
However, some results show that the radio--FIR
correlation holds on 100 pc to
sub-kpc scales \citep{beck88,xu92,hughes06}.
There are some physical
mechanisms that can contribute to
the local variation of the radio--FIR
relation such as dust processing
by SN shocks \citep{cannon05},
dust enrichment by the current star
formation activity \citep{hirashita_hunt08},
amplification of magnetic fields
\citep{beck88,dumas11}, diffusion and escape
of cosmic ray electrons
\citep{murphy06,hughes06,murphy08}, etc.
In particular, the diffusion of cosmic ray electrons
would predict a large FIR-to-radio ratio on small
spatial scales \citep{hughes06}. This is not likely
to be the reason for the II Zw 40 centre though, as long
as its stellar age is too young ($\la 3$ Myr) for
SNRs to generate cosmic ray electrons.


{If young active star-forming regions
generally have similar radio--FIR emission
properties to the central part of II Zw 40,
free--free dominated submm emission and
strong free--free absorption at centimetre and longer
wavelengths
will be a useful guide to find young active
star formation in dense medium. This issue will be
further investigated by collecting a sample of
active star-forming BCDs with submm interferometry
in the future.
}




\section{Conclusion}\label{sec:conclusion}

In order to reveal the radiative properties of
young active starburst, the central star-forming
region in II Zw 40 was observed in the 340~GHz
(880 $\micron$) band at $\sim 5''$ resolution with
SMA. A source associated with the central
star-forming complex was detected with a flux of
$13.6\pm 2.0$ mJy and a 10-arcsec elongation in
the east--west direction. The flux in the central part
of II Zw 40 has been analyzed by
using the theoretical radio SED model developed by
\citet{hirashita06}, and interpreted along with
interferometric measurements
at centimetre wavelengths in the
literature. Then, we have found
{
\begin{enumerate}
\item that the SMA 880 $\micron$
flux is dominated
by free-free emission, and
\item that possible contribution
from dust emission to the SMA flux is less than
$4\pm 2.5$ mJy.
\end{enumerate}
}
Our models have been used to derive the radio--FIR
relation of the II Zw 40 centre.
We have discussed
free--free absorption at low frequencies ($\nu\la$
several GHz;
$\lambda\ga$ several cm) and spatial distribution
of dust ($R_\mathrm{dust}$) as possible factors
affecting the radio--FIR relation.

\section*{Acknowledgments}

We thank Kazushi Sakamoto for his continuous help
for the SMA observation and the data analysis and
his helpful comments on this paper. We thank the
SMA staff for their efforts in running
and maintaining the array.
We are grateful to the
anonymous referee, Takashi Onaka,
and the member of our
star formation group for useful comments that improved
this paper very much. This research
has made use of the
NASA/IPAC Extragalactic Database (NED), which
is operated by the Jet Propulsion Laboratory, California
Institute of Technology, under contract with the National
Aeronautics and Space Administration. This research
is supported through NSC grant 99-2112-M-001-006-MY3.

\bsp

\label{lastpage}

\end{document}